\documentclass[11pt,english,epsf]{scrartcl}
\usepackage{lmodern}
\usepackage[T1]{fontenc}
\usepackage[latin9]{inputenc}
\usepackage{geometry}
\usepackage{amsmath}
\usepackage{amssymb}
\usepackage{graphicx}

\newtheorem{theorem}{Theorem}

\newcommand {\dfn} {\stackrel{\Delta} {=}}
\newcommand {\exe} {\stackrel{\cdot} {=}}
\newcommand {\lexe} {\stackrel{\cdot} {\le}}

\newcommand {\reals} {{\rm I\!R}}

\newcommand {\bx} {\mbox{\boldmath $x$}}
\newcommand {\by} {\mbox{\boldmath $y$}}

\newcommand {\bE} {\mbox{\boldmath $E$}}

\newcommand {\bX} {\mbox{\boldmath $X$}}
\newcommand {\bY} {\mbox{\boldmath $Y$}}

\newcommand{\calC}{{\cal C}}

\newcommand{\calE}{{\cal E}}

\newcommand{\calI}{{\cal I}}

\newcommand{\calT}{{\cal T}}

\newcommand{\calX}{{\cal X}}
\newcommand{\calY}{{\cal Y}}

\allowdisplaybreaks

\begin{document}
\thispagestyle{empty}
\title{Statistical Physics of Random Binning
\thanks{This research was supported by the Israel Science Foundation (ISF),
grant no.\ 412/12.}
}
\author{Neri Merhav}
\date{}
\maketitle

\begin{center}
Department of Electrical Engineering \\
Technion - Israel Institute of Technology \\
Technion City, Haifa 32000, ISRAEL \\
E--mail: {\tt merhav@ee.technion.ac.il}\\
\end{center}
\vspace{1.5\baselineskip}
\setlength{\baselineskip}{1.5\baselineskip}

\begin{center}
{\bf Abstract}
\end{center}
\setlength{\baselineskip}{0.5\baselineskip}
We consider the model of random binning and finite--temperature decoding
for Slepian--Wolf codes, from a statistical--mechanical perspective. While
ordinary random channel coding is intimately related to the random energy
model (REM) -- a statistical--mechanical model
of disordered magnetic materials,
it turns out that random binning (for Slepian--Wolf
coding) is analogous to another, related 
statistical mechanical model of strong disorder, which we call the random
dilution model (RDM). We use the latter analogy to characterize phase transitions
pertaining to finite--temperature Slepian--Wolf decoding, which are somewhat
similar, but not identical, to those of finite--temperature channel decoding.
We then provide the exact random coding exponent of the bit error rate
(BER) as a function of the
coding rate and the decoding temperature, and discuss its properties. 
Finally, a few modifications and extensions of our results are outlined and
discussed.\\

\vspace{0.2cm}

\setlength{\baselineskip}{2\baselineskip}
\newpage

\section{Introduction}

The famous paper by Slepian and Wolf \cite{SW73},
on separate (almost) lossless compression and
joint decompression of statistically dependent
sources, has triggered an intensive
research activity of information theorists 
during the last forty years.
Among its various generalizations and modifications,
several recent works have been dedicated to detailed
performance analysis, first and foremost, to exponential error bounds for the
Slepian--Wolf (SW) decoder.
Specifically, Gallager \cite{Gallager76} obtained a lower bound
on the random coding error exponent associated with
random binning, by employing a very similar technique
to the one he used in his famous derivation of the random (channel) coding error exponent
\cite[Sections 5.5--5.6]{Gallager68}. A few years later, 
this error exponent
was shown by Csisz\'ar, K\"orner
and Marton \cite{CK80}, \cite{CKM77} to be achievable by a universal decoder,
that is independent of the channel.
In \cite{CK81} Csisz\'ar and K\"orner have studied
universally achievable error exponents pertaining to linear codes, as
well as ordinary (non-universal) expurgated exponents.
Later, Csisz\'ar \cite{Csiszar82}
and Oohama and Han \cite{OH94} have developed error exponents for situations
of coded side information.
For high rates at one of the two encoders, Kelly and Wagner \cite{KW11}
have improved relative to these results, but not in the general case.

This paper continues the above described line of work on exponential error bounds associated with random
binning of SW codes, but unlike the previous works mentioned above, this one is
more oriented to the statistical--mechanical
point of view. Specifically, in analogy to the notion of finite--temperature
decoding, originally proposed by Ruj\'an \cite{Rujan93} in the context of
channel coding (see also \cite[Section 6.3.3]{MM09}), 
here we examine a similar finite--temperature decoder for 
SW codes, and analyze it from various aspects. In a nutshell,
finite--temperature decoding amounts to an optimal symbol--error--probability
decoder that is associated with the likelihood function, raised to some power
$\beta\ge 0$, a parameter referred to as the {\it inverse
temperature}, which is a term borrowed from equilibrium statistical
mechanics and the Boltzmann--Gibbs distribution. In channel coding, the motivation for this
parametrization by $\beta$ could either
stem from uncertainty concerning the channel SNR
(which is analogous to the ``temperature'' of the real channel), or from
the fact that it enables to analyze both the optimal
symbol--error--probability decoder and the 
optimal block--error--probability decoder on the same footing
($\beta=1$ and $\beta\to\infty$, respectively \cite[p.\ 118]{MM09}).
In SW coding, the motivations are similar.

Focusing mostly on the ``one--sided'' version of the Slepian--Wolf setting,
where one source is compressed while the other one is available (e.g.,
after perfect reconstruction) as side information at the decoder, we derive several results
in this paper. 
First, we present a statistical--mechanical model, henceforth referred to as
the {\it random dilution model} (RDM), which is a natural analogue of 
the random binning mechanism, exactly in the same way that the random energy model
(REM) of statistical mechanics is the physical analogue of the random coding mechanism (see
also \cite{fnt}, \cite[Chapters 5 and 6]{MM09} and references therein).
The RDM was already mentioned briefly in an earlier work \cite{swlist}, but
was not developed in detail therein. 

Secondly, in analogy to the phase diagrams of finite--temperature channel coding,
provided in \cite{REM1st} and \cite[Sect.\ 6.3.3]{MM09}, here we
provide a statistical--mechanical characterization in the form of a {\it phase
diagram} of SW codes
with random binning, in the plane of $R$ vs.\ $T$, where $R$ is the coding
rate and $T=1/\beta$ is the decoding temperature. As in channel coding, there
are three different phases, in which the kinds of behavior of the posterior
distribution of the source given the side information and the bin index, are
completely different. We will elaborate on these phases in the sequel.
Generally speaking,
the phase diagram of a finite--temperature SW decoder appears similar to the
``mirror image'' of the one of channel coding, where the axis of the rate $R$ is
flipped over. On the one hand, this seems to make sense, 
in view of the fact that a SW code at rate $R$ is
nearly equivalent to
a channel code at rate $H-R$, where $H$ is the entropy of the compressed
source. However, this equivalence is not perfect, as there are also some
non--trivial differences
between channel coding and SW coding. Accordingly, there are differences also in
the phase diagram beyond the aforementioned ``mirror reflection''.

Next, we derive the exact exponent of the symbol error probability of the finite--temperature
decoder, as a function of $R$ and $\beta$, and we make a few observations
concerning the properties of the resulting error exponent, denoted $E(R,\beta)$. It
turns out that $E(R,\beta)$ also exhibits phase transitions, which are related
to the above mentioned phases of the posterior, but are not identical.

Finally, we outline a few extensions and modifications of the above described
results, in several directions, including: mismatched decoding,
universal decoding, variable--rate coding, and the ``two--sided'' SW problem,
where both sources are compressed separately and decompressed jointly.

The outline of the paper is as follows. 
In Section 2, we establish notation conventions, define the problem setting,
and provide some physics background.
In Section 3, we derive the phase diagram, and in
in Section 4, we derive the error exponent of the finite--temperature decoder.

\section{Notation Conventions, Problem Setting and Background}

\subsection{Notation Conventions}

Throughout the paper, random variables will be denoted by capital
letters, specific values they may take will be denoted by the
corresponding lower case letters, and their alphabets
will be denoted by calligraphic letters. Random
vectors and their realizations will be denoted,
respectively, by capital letters and the corresponding lower case letters,
both in the bold face font. Their alphabets will be superscripted by their
dimensions. For
example, the random vector $\bX=(X_1,\ldots,X_N)$, ($N$ -- positive
integer) may take a specific vector value $\bx=(x_1,\ldots,x_N)$
in $\calX^N$, the $N$--th order Cartesian power of $\calX$, which is
the alphabet of each component of this vector.

The expectation operator will be denoted by $\bE\{\cdot\}$.
Logarithms and exponents will be understood to be taken to the natural base
unless specified otherwise.
The indicator function will be denoted by $\calI(\cdot)$. The notation
function $[t]_+$ will be defined as $\max\{t,0\}$. For two positive sequences,
$\{a_N\}$ and $\{b_N\}$, the notation $a_N\exe b_N$ will mean asymptotic
equivalence in the exponential scale, that is,
$\lim_{N\to\infty}\frac{1}{N}\log(\frac{a_N}{b_N})=0$.
Similarly, $a_N\lexe b_N$ will mean
$\limsup_{N\to\infty}\frac{1}{N}\log(\frac{a_N}{b_N})\le 0$, and so on.

\subsection{Problem Setting and Objectives}

Let $\{(X_i,Y_i)\}_{i=1}^N$ be $N$ independent copies of a random vector
$(X,Y)$, distributed according to a given probability mass function
$P(x,y)$, where $x$ and $y$ take on values
in finite alphabets, $\calX$ and $\calY$, respectively. The source vector
$\bx=(x_1,\ldots,x_N)$, which is a generic realization of
$\bX=(X_1,\ldots,X_N)$,
is compressed at the encoder by random binning, that
is, each $N$--tuple $\bx\in\calX^N$ is randomly
and independently assigned to one out of $M=e^{NR}$ bins,
where $R$ is the coding rate in nats per symbol.
Given a realization of the random partitioning into bins (revealed to both the
encoder and
the decoder), let
$f:\calX^N\to\{0,1,\ldots,M-1\}$ denote the encoding function, i.e.,
$u=f(\bx)$
is the encoder output.
Accordingly, the inverse image of $u$, defined as
$f^{-1}(u)=\{\bx:~f(\bx)=u\}$, is the bin of all source vectors mapped by the
encoder into
$u$. The decoder has access to $u$ and to $\by=(y_1,\ldots,y_N)$, which is a
realization of $\bY=(Y_1,\ldots,Y_N)$, namely, the side information at the decoder.

As is well known, the optimal decoder in the sense of minimum word error probability is
the word--level maximum a-posteriori (MAP) decoder,
\begin{equation}
\label{block}
\hat{\bx}=\mbox{arg}\max_{\bx\in f^{-1}(u)} P(\bx|\by)
=\mbox{arg}\max_{\bx\in f^{-1}(u)} P(\bx,\by).
\end{equation}
Similarly, the optimal decoder in the sense of minimum symbol error probability is
given by the symbol--level MAP decoder,
\begin{equation}
\label{symbol}
\hat{x}_i=\mbox{arg}\max_{x\in\calX}\sum_{\bx\in f^{-1}(u):~x_i=x}
P(\bx,\by),~~~~~~i=1,2,\ldots,N.
\end{equation}
Following the notion of finite--temperature decoding in channel coding \cite{Rujan93} (see also
\cite[Section 6.3.3]{MM09}), we consider a parametric family of decoders, that
generalizes both (\ref{block}) and (\ref{symbol}), and which is of
the form
\begin{equation}
\label{ft}
\hat{x}_i=\mbox{arg}\max_{x\in\calX}\sum_{\bx\in f^{-1}(u):~x_i=x} P^\beta (\bx,\by),
~~~~~~i=1,2,\ldots,N,
\end{equation}
where the parameter $\beta\ge 0$ is referred to as the {\it inverse
temperature}, a term borrowed from equilibrium statistical physics (see next
subsection).
The motivation for considering the finite--temperature decoder is two--fold
(see also \cite{REM1st} for even more motivations): 
First, as said, it is a common generalization of both the
symbol--level MAP decoder ($\beta=1$) and the
word--level MAP decoder ($\beta\to\infty$). Secondly, in some important cases, it refers to a situation
of a certain mismatch that may stem from uncertainty concerning 
the joint distribution of $(X,Y)$, or even more specifically,
the quality of the `channel' $P(y|x)$, connecting $X$
to $Y$. For example, if $(X,Y)$ is a double binary symmetric source (BSS),
that is, $X$ is a BSS and $Y$ given $X$ is generated by binary symmetric channel
(BSC), then the choice of $\beta$ manifests the decoder's `belief' concerning the
quality of this BSC: $\beta < 1$ corresponds to a pessimistic decoder,
whereas $\beta > 1$ is associated with an optimistic one.

In order to understand the behavior of the finite--temperature decoder
(\ref{ft}), one has to first gain some insight concerning the behavior the
posterior distribution induced by that decoder, namely, 
\begin{equation}
\label{ftposterior}
P_\beta(\bx|\by,u)=\left\{\begin{array}{ll}
\frac{P^\beta(\bx,\by)}{\sum_{\bx'\in f^{-1}(u)}P^\beta(\bx',\by)} & \bx\in
f^{-1}(u)\\
0 & \mbox{elsewhere}\end{array}\right.
\end{equation}
Accordingly, we first focus on studying the properties
of this posterior in the random binning regime. In particular, similarly as
in \cite{REM1st} and \cite{MM09}, we view it as
an instance of the Boltzmann--Gibbs (B-G) distribution of statistical
mechanics by rewriting it in the form
\begin{equation}
\label{bg}
P_\beta(\bx|\by,u)=\left\{\begin{array}{ll}
\frac{\exp\{-\beta \calE(\bx,\by)\}}{\sum_{\bx'\in f^{-1}(u)}\exp\{-\beta
\calE(\bx',\by)\}} & \bx\in
f^{-1}(u)\\
0 & \mbox{elsewhere}\end{array}\right.
\end{equation}
where $\calE(\bx,\by)$ is the energy function (Hamiltonian), defined as
$\calE(\bx,\by)\dfn-\ln P(\bx,\by)$. We then study the phase diagram of 
the corresponding statistical--mechanical model in the plane of $R$ vs.\
$\beta$, or more precisely, $R$ vs.\ $T$, where $T=1/\beta$ is the decoding
temperature.
In analogy to the role of the random
energy model (REM) as the statistical--mechanical counterpart of ordinary
random coding (see 
\cite[Chap.\ 6]{fnt}, 
\cite[Chapters 5 and 6]{MM09}), 
it turns out that random
binning corresponds to a somewhat different (though related) physical model,
which we call the random dilution model (RDM), and which was first mentioned in
\cite{swlist}. The RDM and its relevance to random binning will be
presented in the next subsection.

Our second objective would be to derive the {\it exact} error exponent of the
symbol error probability, denoted
$E(R,\beta)$, that is associated with the
finite--temperature decoder (\ref{ft}), as a function of $R$ and $\beta$, in the
random binning regime. The properties of $E(R,\beta)$, as well as its phase
diagram, will be studied in some detail. 

Finally, we briefly discuss several variations of these results, covering
situations of mismatch, universal decoding, variable--rate compression, and
the ``two--sided'' version of SW coding, namely, separate encodings and 
joint decoding of both sources (as opposed to the ``one--sided'' version 
described above, of encoding and decoding of one source while the
other one serves as side information at the decoder).

\subsection{Background: The REM, the RDM and Random Binning}

In ordinary random coding, the analysis of
bounds on the probability of error (especially in
Gallager's method) are often associated with expressions of the form
$\sum_{\bx\in\calC} P^\beta(\by|\bx)$, where $P(\by|\bx)$ is the
conditional distribution pertaining to the channel, $\calC$ is randomly drawn
codebook and $\beta > 0$ is a parameter. 
As described in \cite[Chap.\ 6]{fnt}, from the statistical--mechanical
perspective,  this can be thought of as 
a partition function (depending on $\by$):
\begin{equation}
Z(\beta|\by)=\sum_{\bx\in\calC} e^{-\beta \calE(\bx,\by)},
\end{equation}
where $\beta$ is the inverse temperature
and where here the energy function
is $\calE(\bx,\by)=-\ln P(\by|\bx)$. 
The same partition function is relevant for the Boltzmann--Gibbs form of the
finite--temperature posterior associated with the channel decoder, in the
spirit of eqs.\ (\ref{ftposterior}) and (\ref{bg}):
\begin{eqnarray}
P_\beta(\bx|\by)&=&\left\{\begin{array}{ll}
\frac{P^\beta(\by|\bx)}{Z(\beta|\by)} & \bx\in
\calC\\
0 & \mbox{elsewhere}\end{array}\right.\\
&=&\left\{\begin{array}{ll}
\frac{\exp\{-\beta \calE(\bx,\by)\}}{Z(\beta|\by)} & \bx\in
\calC\\
0 & \mbox{elsewhere}\end{array}\right.
\end{eqnarray}
As the codewords are selected
independently at
random, then for fixed $\by$, the energy values $\{\calE(\bx,\by),~\bx\in\calC\}$ are
i.i.d.\ random variables. This is essentially the same as in the {\it random
energy model} (REM), a well known model of disorder in statistical physics of
spin glasses, which undergoes a phase transition: below a certain
temperature ($\beta > \beta_c$), the system is frozen in the sense that
the partition function is dominated by a non--exponential number of
microstates $\{\bx\}$ at the ground--state energy (zero thermodynamical entropy).
This is called the
{\it frozen phase} or the {\it glassy phase}.
The other phase, $\beta < \beta_c$, is called the
{\it paramagnetic phase}
(see more details in \cite[Chap.\
5]{MM09}). Owing to this analogy between the REM and random coding, 
the corresponding exponential error bounds associated
with random coding then undergo a similar phase transition (see \cite{fnt}
and references therein).

In random binning, as opposed ordinary random coding, the mechanism 
is somewhat different, and
the analogous statistical--mechanical model, which we call the
RDM, is defined as follows.
Consider a partition function of a certain physical system,
with a microstate $\bx$ and Hamiltonian $\calE(\bx)$, i.e.,
\begin{equation}
Z(\beta)=\sum_{\bx\in\calX^N} e^{-\beta \calE(\bx)},~~~\beta > 0,
\end{equation}
where $\beta$ is, as said, the inverse temperature.
The {\it diluted} version of $Z(\beta)$, according to the RDM (hence the name), is defined as
\begin{equation}
Z_{\mbox{\tiny D}}(\beta)=\sum_{\bx\in\calX^N} I(\bx)\cdot e^{-\beta \calE(\bx)},
\end{equation}
where $\{I(\bx),~\bx\in\calX^N\}$
are i.i.d.\ Bernoulli random variables with
$p\dfn \mbox{Pr}\{I(\bx)=1\}=1-\mbox{Pr}\{I(\bx)=0\}=e^{-NR}$ for all
$\bx\in\calX^N$, and $R\ge 0$ is a given constant.
Thus, $Z_{\mbox{\tiny D}}(\beta)$ is a partial version of the full partition
function $Z(\beta)$, with randomly chosen (surviving) microstates $\{\bx\}$. 
Equivalently, $Z_{\mbox{\tiny D}}(\beta)$ can be thought of as being 
defined just like $Z(\beta)$, but
with an Hamiltonian redefined as $\calE_{\mbox{\tiny
D}}(\bx)=\calE(\bx)+\psi(\bx)$, where
$\psi(\bx)$ are i.i.d.\ random variables, taking the value $\psi(\bx)=0$ with probability
$e^{-NR}$ and the value $\psi(\bx)=\infty$ with probability $1-e^{-NR}$.
From the physical point of view, $\psi(\bx)$ can be thought of as
some disordered potential energy function,
that due to long--range interactions, disables
access to certain points in the configuration space (those that have
not `survived' the dilution).

Let $s(\epsilon)$ denote the normalized
(per--particle)
entropy as a function of the normalized energy, associated with the full system, $Z(\beta)$.
More precisely, denoting by $\Omega_N(E)$ the number of vectors $\{\bx\}$
for which $\calE(\bx)=E$, then
\begin{equation}
\label{entropydef}
s(\epsilon)\dfn\lim_{N\to\infty}\frac{\ln\Omega_N(N\epsilon)}{N}, 
\end{equation}
provided
that the limit exists.
Let $\Delta \epsilon$ be an arbitrarily small quantity (increment) of the normalized energy. Then,
\begin{equation}
Z_{\mbox{\tiny D}}(\beta)\approx\sum_i \left[\sum_{\bx:~Ni\Delta \epsilon\le
\calE(\bx)<N(i+1)\Delta \epsilon}I(\bx)\right]\cdot e^{-\beta Ni\Delta \epsilon}.
\end{equation}
Observe that the expression in the square brackets
is a binomial RV with exponentially about $e^{Ns(i\Delta \epsilon)}$ trials and
probability of success $p=e^{-NR}$. Thus, in a typical realization
the RDM, this number is about $\exp\{N[s(i\Delta \epsilon)-R]\}$ whenever
$s(i\Delta \epsilon)-R\ge 0$ and by zero otherwise. Thus, in the limit of $\Delta
\epsilon\to 0$,
\begin{eqnarray}
\phi_{\mbox{\tiny D}}(\beta)=\lim_{N\to\infty}\frac{\ln Z_{\mbox{\tiny D}}(\beta)}{N}&=&
\sup_{\{\epsilon:~s(\epsilon)\ge R\}}[s(\epsilon)-R-\beta \epsilon]\\
&=&\left\{\begin{array}{ll}
\phi(\beta)-R & \beta < \beta_c\\
-\beta \epsilon_0 & \beta \ge \beta_c\end{array}\right.
\end{eqnarray}
where $\phi(\beta)$ is the asymptotic normalized log--partition function
associated with the full system (before the dilution),
$\epsilon_0$ is ground--state (minimum energy state) 
of the RDM, i.e., the solution to the equation $s(\epsilon)=R$, and
$\beta_c=s'(\epsilon_0)$, $s'(\cdot)$ being the derivative of $s(\cdot)$. 
Note that both $\beta_c$ and $\epsilon_0$ depend on $R$.
Thus, the system undergoes a glassy phase transition (similarly to the REM) at
inverse temperature $\beta_c$. At low temperatures, it freezes at zero
entropy and $Z_{\mbox{\tiny D}}(\beta)$ is dominated by a subexponential
number of configurations $\{\bx\}$ at the ground--state energy
$E_0=N\epsilon_0$. This is then the glassy phase of the RDM.
Above the critical temperature (the 
paramagnetic phase), the diluted system behaves essentially like the full
one, i.e., it is dominated by exponentially many
(actually, about
$e^{N[s(\epsilon_\beta)-R]}$) configurations with energy
$E_\beta=N\epsilon_\beta$, where $\epsilon_\beta$ is
the typical per--particle energy level pertaining to inverse temperature $\beta$ in the full
system, that is, the solution to the equation $s'(\epsilon)=\beta$.
Note that 
\begin{equation}
\beta_c(R)=s'[s^{-1}(R)], 
\end{equation}
where $s^{-1}(\cdot)$ is the inverse
function of $s(\cdot)$ (in the range where it is monotonically increasing and
hence one--to--one). In this range $s^{-1}(\cdot)$ is increasing as well, but since
$s'$ is decreasing, then $\beta_c$ is a decreasing function of $R$. This means
that, as the
dilution becomes more aggressive, the critical temperature goes up.
Since $s(\cdot)$ is concave (see, e.g., \cite[pp.\ 13--14]{fnt}) 
and increasing, $s^{-1}(\cdot)$ is convex, so the overall behavior
depends on $s'(\cdot)$. 

\vspace{0.1cm}

\noindent
{\it Example 1.} Let $\calE(\bx)=\frac{\kappa}{2}\|\bx\|^2$, with
$\calX=\{0,\pm a,\pm 2a,\ldots\}$, i.e., an harmonic potential applied to
particles in a grid with spacing $a$. Then, $\Omega_N(E)$ is approximately the
volume of the shell of a hyper--sphere of radius $\sqrt{2E/\kappa}$, divided by an elementary
volume of the grid cube $a^N$, which yields
\begin{equation}
s(\epsilon)=\frac{1}{2}\ln\frac{4\pi
e\epsilon}{\kappa a^2},
\end{equation}
and so,
\begin{equation}
s'(\epsilon)=\frac{1}{2\epsilon}.
\end{equation}
and
\begin{equation}
s^{-1}(R)=\frac{\kappa a^2}{4\pi e}\cdot e^{2R}.
\end{equation}
Thus,
\begin{equation}
\beta_c(R)=\frac{2\pi e}{\kappa a^2}\cdot e^{-2R},
\end{equation}
meaning that the critical temperature grows exponentially with $R$.
This concludes Example 1.

\vspace{0.1cm}

\noindent
{\it Remark 1.} A slightly more general version of the RDM replaces the fixed
parameter $R$ by a function of $\epsilon$, that is, $p=e^{-NR(\epsilon)}$. The
analysis is essentially the same as before, except that now, the range of maximization
$\{\epsilon:~s(\epsilon) \ge R\}$ is replaced by
$\{\epsilon:~s(\epsilon) \ge R(\epsilon)\}$, which, depending on the form of
the function $R(\cdot)$, might be rather different.

Finally, to see the relevance of the RDM to random binning for SW coding,
let us return to the problem setting
described in Subsection 2.2, and consider the partition function
\begin{equation}
\label{Z}
Z(\beta|\by,u)=\sum_{\bx\in f^{-1}(u)}P^\beta(\bx,\by)=
\sum_{\bx\in\calX^N}\exp\{-\beta \calE(\bx,\by)\}\cdot
\calI[f(\bx)=u],
\end{equation}
pertaining to the Boltzmann-Gibbs distribution (\ref{bg}).
We can think
of this as an instance of the RDM, with $I(\bx)=\calI[f(\bx)=u]$, i.e., the
microstates $\{\bx\}$ that `survive' the dilution are only those for which the
randomly selected bin index happens to coincide with the given $u$, 
which is the case with probability $e^{-NR}$, exactly
like in the above defined RDM.

\section{Phase Diagram of the Finite--Temperature Posterior}

In this subsection, we characterize the phase diagram 
of the partition function (\ref{Z}), pertaining to
the finite--teperature posterio (\ref{bg}), for a typical realization of
the random binning scheme and a typical realization of $(\bX,\bY)$.
Generally speaking, this derivation is in the spirit of those in 
\cite[Chap.\ 6]{fnt} and
\cite[Section 6.3.3]{MM09},
but there are some important differences.

We begin by decomposing
$Z(\beta|\by,u)$ as 
\begin{equation}
Z(\beta|\by,u)=Z_{\mbox{\tiny c}}(\beta|\by,u)+Z_{\mbox{\tiny e}}(\beta|\by,u),
\end{equation}
where $Z_{\mbox{\tiny c}}(\beta|\by,u)=e^{-\beta \calE(\bx,\by)}$ is the contribution
of the {\it correct} $\bx$ that was actually emitted by the source, 
whereas $Z_{\mbox{\tiny e}}(\beta|\by,u)$ is the sum of
contributions of all other source vectors. For a typical realization
$(\bx,\by)$ of
$(\bX,\bY)$, $\calE(\bx,\by)$ is about $NH(X,Y)$ (by the weak law of large
numbers), and so,
$Z_{\mbox{\tiny c}}(\beta|\by,u)$ is about $e^{-\beta N H(X,Y)}$,
where $H(X,Y)$ is the joint entropy of $(X,Y)$. What makes the real $\bx$
emitted having a special stature here is the fact that it surely survives the
dilution, as
$u$ is, by definition, the bin index of $\bx$.

We next address the behavior of the second term, $Z_{\mbox{\tiny
e}}(\beta|\by,u)$. To this end, we need the entropy function $s(\epsilon)$
(see (\ref{entropydef})) of 
the full (non--diluted) system.
Using the method of types, it is easily seen
that for a typical $\by$, this function
is given by
\begin{equation}
s(\epsilon)=\max_{\{Q(x|y):~-\sum_{x,y}P(y)Q(x|y)\ln
P(x,y)=\epsilon\}}\sum_yP(y)\sum_xQ(x|y)\ln\frac{1}{Q(x|y)}.
\end{equation}
It would be instructive and useful to
characterize the form of the optimal `channel' $\{Q(x|y)\}$, call it $\{Q^*(x|y)\}$,
that achieves $s(\epsilon)$.
Intuitively, $s(\epsilon)$ can be thought of as the overall
per--particle entropy of a mixture of systems, indexed by $y$, each one with
$NP(y)$ particles and Hamiltonian $\calE(x,y)=-\ln P(x,y)$, where $x$ plays the
role of a micro-state and $y$ is the index. In thermal equilibrium, all systems
are at the same temperature, which we will denoted by $\tau=1/\alpha$, and the Boltzmann factor is
proportional to $e^{-\alpha \calE(x,y)}=P^\alpha(x,y)$, where $\alpha$ is chosen
so as to meet the constraint.

More precisely,
let $\zeta(\alpha|y)=\sum_x P^\alpha(x,y)$,
$\alpha\in\reals$.
We argue that the conditional distribution $Q^*(x|y)$ that achieves
$s(\epsilon)$ is always of the form
\begin{equation}
Q^*(x|y)=Q_\alpha(x|y)\dfn\frac{P^\alpha(x,y)}{\zeta(\alpha|y)},
\end{equation}
where $\alpha$ is chosen to satisfy the constraint
\begin{equation}
\label{alphaeps}
\sum_{x,y}P(y)Q_\alpha(x|y)\calE(x,y)=
-\sum_{x,y}P(y)Q_\alpha(x|y)\ln P(x,y)=\epsilon.
\end{equation}
Note that for $\alpha\to\infty$, $Q_\alpha(x|y)$ tends to put all its mass on
the letter $x$ which maximizes $P(x,y)$ and the resulting energy is
$\epsilon_{\min}=\sum_yP(y)\min_x \ln[1/P(x,y)]$, whereas
for $\alpha\to-\infty$, $Q_\alpha(x|y)$ tends to put all its mass on
the letter $x$ which minimizes $P(x,y)$ and the resulting energy is
$\epsilon_{\max}=\sum_yP(y)\max_x \ln[1/P(x,y)]$. Thus, as $\alpha$ exhausts
the real line, the entire energy range $(\epsilon_{\min},\epsilon_{\max})$ is
covered. The optimality of $Q_\alpha(x|y)$ follows from the 
following consideration:
\begin{eqnarray}
\label{Qoptimality}
0&\le&\sum_{x,y}P(y)Q(x|y)\ln\frac{Q(x|y)}{Q_\alpha(x|y)}\nonumber\\
&=&
\sum_{x,y}P(y)Q(x|y)\ln\frac{Q(x|y)\zeta(\alpha|y)}{P^\alpha(x,y)}\nonumber\\
&=&\sum_yP(y)\ln\zeta(\alpha|y)+\alpha\epsilon+\sum_yP(y)\sum_xQ(x|y)\ln
Q(x|y)
\end{eqnarray}
or
\begin{equation}
\sum_yP(y)\sum_xQ(x|y)\ln\frac{1}{Q(x|y)}\le \sum_yP(y)\ln\zeta(\alpha|y)+\alpha\epsilon,
\end{equation}
with equality for $Q(x|y)=Q_\alpha(x|y)$.
It is also seen that
\begin{equation}
s(\epsilon)=\sum_yP(y)\ln\zeta(\alpha|y)+\alpha\epsilon,
\end{equation}
where it should be kept in mind that $\alpha$ is itself a function of
$\epsilon$, defined by the constraint (\ref{alphaeps}).

\begin{figure}[ht]
\hspace*{2cm}\input{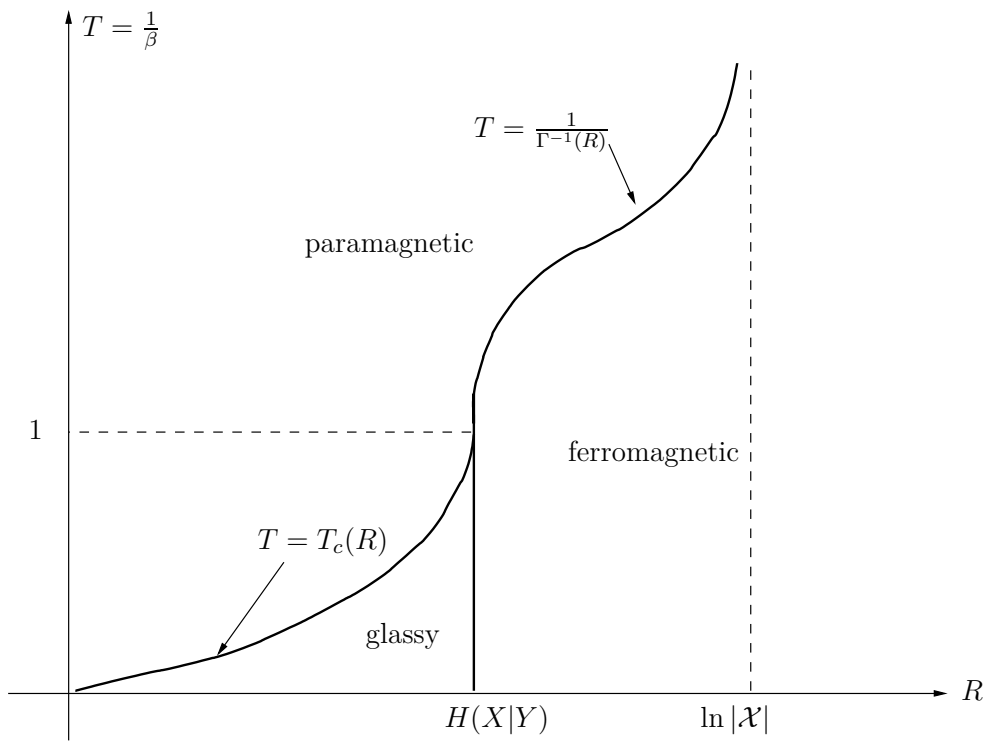}
\caption{\small Phase diagram of $Z(\beta|\by)$ (for a typical $\by$) 
in the plane of the decoding temperature $T$ vs.\
the coding rate $R$.}
\label{fig1}
\end{figure}

Now, as $Z_{\mbox{\tiny e}}(\beta|\by,u)$ is associated with the RDM, it 
has two phases: the paramagnetic phase and the glassy phase.
in the plane of $\beta$ vs.\ $R$, where
the boundary is $\beta=\beta_c(R)$ with the entropy function $s(\cdot)$ as above.
The contribution of $Z_{\mbox{\tiny c}}(\beta|\by,u)$ introduces a third phase -- the
so called, {\it ordered phase} or {\it ferromagnetic phase}. 
The ferromagnetic phase dominates the glassy phase when
$\beta H(X,Y)\le \beta s^{-1}(R)$, namely, when $R \ge s[H(X,Y)]$.
Now, we argue that $s[H(X,Y)]=H(X|Y)$. 
To prove this, first observe that obviously, $s[H(X,Y)]\ge
H(X|Y)$ (by choosing $Q(x|y)=P(x|y)$). On the other hand, the reversed
inequality is obtained by repeating eq.\ (\ref{Qoptimality}) with the choice $\alpha=1$.
Thus, the boundary between the ferromagnetic phase and the glassy phase is
given by $R=H(X|Y)$ (the vertical line in the phase diagram of Fig.\
\ref{fig1}).
Note also that $\beta_c[H(X|Y)]=1$.

Now, the normalized log--partition function of the non--diluted system,
$\phi(\beta)$, is
obtained by
\begin{eqnarray}
e^{N\phi(\beta)}&=&\sum_{\bx} P^\beta(\bx,\by)\\
&=&\sum_{\bx} \prod_{i=1}^N P^\beta(x_i,y_i)\\
&=&\prod_{i=1}^N \left[\sum_xP^\beta(x,y_i)\right]\\
&=&\prod_{y\in\calY} \left[\sum_xP^\beta(x,y)\right]^{NP(y)}\\
&=&\exp\left\{N\sum_{y\in\calY}
P(y)\ln\left[\sum_xP^\beta(x,y)\right]\right\},
\end{eqnarray}
i.e.,
\begin{equation}
\phi(\beta)=\sum_{y\in\calY}
P(y)\ln\left[\sum_xP^\beta(x,y)\right].
\end{equation}
It follows that for the ferromagnetic component to dominate also the paramagnetic
component, we must have
\begin{equation}
\beta H(X,Y)\le -\sum_{y\in\calY} P(y)\ln\left[\sum_xP^\beta(x,y)\right]+R,
\end{equation}
or, equivalently,
\begin{equation}
R\ge \beta H(X,Y)+\sum_{y\in\calY}
P(y)\ln\left[\sum_xP^\beta(x,y)\right]\dfn \Gamma(\beta),
\end{equation}
and so the boundary is given by $R=\Gamma(\beta)$ or $T=1/\Gamma^{-1}(R)$.
The boundary between the glassy phase and the paramagnetic phase is, of
course, $\beta=\beta_c(R)$ or $T=T_c(R)\dfn 1/\beta_c(R)$, 
as mentioned already in the general discussion
on the RDM.
The phase diagram of finite--temperature random binning 
appears in Fig.\ \ref{fig1}, as a partition of the plane of $T=1/\beta$ vs.\
$R$ into the three regions mentioned.
As can be seen, qualitatively speaking, it looks
quite like the mirror image of the phase diagram of random coding for channels
\cite{REM1st}, \cite[p.\ 119, Fig.\ 6.5]{MM09}, since a
rate $R$ SW code essentially operates like a channel code at rate $H(X)-R$. However, the
equations of the boundary curves $T=T_c(R)$ and $T=1/\Gamma^{-1}(R)$ here and
in channel coding are completely different due to several reasons: 
\begin{enumerate}
\item In SW
coding, the typical size of a bin (which is analogous to the size of the corresponding
channel codebook) is a random variable, which fluctuates around
$|\calX|^N\cdot e^{-NR}$. Only about $\exp\{N[H(X)-R]\}$ members of this bin
are typical to the source, but when error exponents and large deviations
effects are considered, the a-typical bin members may also play a non--trivial
role.
\item Unlike in traditional channel coding, the prior of the input is
not necessarily uniform across the bin, as it depends on the source (just like in
joint source--channel coding).
\item The compositions (types) of
the codewords are random. 
\end{enumerate}
All these differences make the analogy between SW coding and channel coding
rather non--trivial in our context.
A few comments are in order concerning possible extensions and modifications
of the above phase diagram.

\noindent
1. {\it Variable--rate SW codes.} Variable--rate SW codes 
may be related to the generalized version of the
RDM that allows $R$ to be energy--dependent (see Remark 1). In the context of
variable--rate SW coding, this requires a slight modification, as 
$R$ may be allowed to depend only on $\bx$, but not on $\by$. A
plausible approach is then to let $R$ depend on $\bx$ only via the type class of
$\bx$ (see \cite{nirsw}).

\vspace{0.1cm}

\noindent
2. {\it Mismatch.} The above analysis can easily be extended to a situation of a general mismatch.
Suppose that the partition function is defined in terms of a mismatched
model $\tilde{P}(\bx,\by)$, where we assume, without loss of generality, that 
$\tilde{P}(y)=P(y)$, because as far as the finite--temperature decoder is concerned, 
a general $\tilde{P}(\bx,\by)$ is equivalent to $P(\by)\tilde{P}(\bx|\by)$,
where $\tilde{P}(\bx|\by)$ is the conditional distribution induced by
$\tilde{P}(\bx,\by)$.
Accordingly, in Fig.\ \ref{fig1}, the ferromagnetic--glassy boundary would be
replaced by the vertical straight line
$$R=-\bE\ln\tilde{P}(X|Y)=-\sum_{x,y}P(x,y)\ln\tilde{P}(x|y),$$
the ferromagnetic--paramagnetic boundary would be modified to
$$R=\tilde{\Gamma}(\beta)=-\beta\bE\ln\tilde{P}(X,Y)+
\sum_y P(y)\ln\left[\sum_x\tilde{P}^\beta(x,y)\right],$$
and the paramagnetic--glassy boundary would become
$$\beta=\tilde{\beta}_c(R)=\tilde{s}'[\tilde{s}^{-1}(R)],$$
where $\tilde{s}(\epsilon)$ is defined similarly as $s(\epsilon)$, except
that $P(x,y)$ is replaced by $\tilde{P}(x,y)$.

\vspace{0.1cm}

\begin{figure}[ht]
\hspace*{2cm}\input{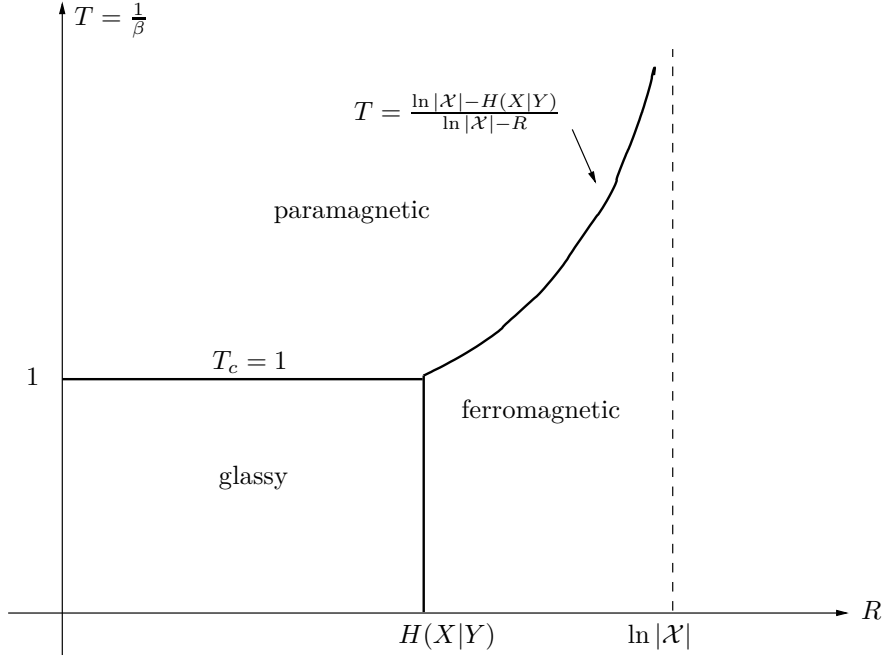}
\caption{\small Phase diagram of the finite--temperature minimum entropy
decoder in the plane of the decoding
temperature $T$ vs.\
the coding rate $R$.}
\label{fig1u}
\end{figure}

\noindent
3. {\it Universal decoding.} 
It is interesting to analyze similarly the partition function pertaining to
a finite--temperature version of the (universal) minimum conditional entropy decoder.
The only difference is that here, the Hamiltonian is replaced by 
$\calE(\bx,\by)=-N\sum_yP(y)\sum_xQ(x|y)\ln Q(x|y)$, where $Q(x|y)$ is the
conditional empirical distribution of $X$ given $Y$, induced from $(\bx,\by)$. 
Obviously, in this
case, $s(\epsilon)=\epsilon$, and so, for a typical $\by$:
\begin{eqnarray}
\lim_{N\to\infty}\frac{\ln Z_e(\beta|\by)}{N}&=&
\sup_{\{\epsilon:~\epsilon\ge R\}}[\epsilon(1-\beta)-R]\\
&=&\left\{\begin{array}{ll}
(1-\beta)\ln|X|-R & \beta < 1\\
-\beta R & \beta \ge 1\end{array}\right.
\end{eqnarray}
which means that the critical temperature is always $T_c=1$, independently of
$R$. Thus, the paramagnetic--glassy boundary becomes the horizontal straight
line $T_c=1$. The paramagnetic--ferromagnetic boundary is now 
$$R=(1-\beta)\ln|\calX|+\beta H(X|Y),$$
or equivalently,
$$T=\frac{\ln|\calX|-H(X|Y)}{\ln|\calX|-R}.$$
The phase diagram is depicted in Fig.\ \ref{fig1u}.
As can be seen, the price of the universality is that the paramagnetic phase
partly `invades' into the previous area of the ferromagnetic phase, and
that, similarly, the glassy phase has
expanded at the expense of the paramagnetic phase.

4. {\it Two--Sided SW Coding.}
When both $\bx$ and $\by$ are encoded, at rates $R_X$ and $R_Y$, respectively,
the partition function becomes
\begin{equation}
Z(\beta|u,v)=\sum_{\bx',\by'}P^\beta(\bx',\by')\cdot I[f_X(\bx')=u]\cdot
I[f_Y(\by')=v].
\end{equation}
Here, we should distinguish between four terms:
\begin{eqnarray}
Z_{\mbox{\tiny cc}}(\beta|u,v)&=&P^\beta(\bx,\by)=e^{-\beta\ln[1/P(\bx,\by)]}\nonumber\\
Z_{\mbox{\tiny ec}}(\beta|\by,u,v)&=&\sum_{\bx'\ne\bx}P^\beta(\bx',\by)\cdot
I[f_X(\bx')=u]\\
Z_{\mbox{\tiny ce}}(\beta|\bx,u,v)&=&\sum_{\by'\ne\by}P^\beta(\bx,\by')\cdot I[f_Y(\by')=v]\\
Z_{\mbox{\tiny ee}}(\beta|u,v)&=&\sum_{\bx'\ne\bx,\by'\ne\by}P^\beta(\bx',\by')\cdot I[f_X(\bx')=u]\cdot
I[f_Y(\by')=v].
\end{eqnarray}
As before, $Z_{\mbox{\tiny cc}}(\beta|u,v)$ is typically about $e^{-N\beta H(X,Y)}$.
$Z_{\mbox{\tiny ec}}(\beta|\by,u,v)$ is exactly the same as the earlier
$Z_{\mbox{\tiny e}}(\beta|\by,u)$,
and so is, $Z_{\mbox{\tiny ce}}(\beta|\bx,u,v)$, with the roles of $\bx$ and $\by$
being interchanged. Thus, we define
\begin{equation}
s_{X|Y}(\epsilon)=\max_{\{Q(x|y):~-\sum_{x,y}P(y)Q(x|y)\ln
P(x,y)=\epsilon\}}\sum_{x,y}P(y)Q(x|y)\ln\frac{1}{Q(x|y)}
\end{equation}
\begin{equation}
s_{Y|X}(\epsilon)=\max_{\{Q(y|x):~-\sum_{x,y}P(x)Q(y|x)\ln
P(x,y)=\epsilon\}}\sum_{x,y}P(x)Q(y|x)\ln\frac{1}{Q(y|x)}
\end{equation}
and
\begin{equation}
s_{XY}(\epsilon)=\max_{\{Q(x,y):~-\sum_{x,y}Q(x,y)\ln
P(x,y)=\epsilon\}}\sum_{x,y}Q(x,y)\ln\frac{1}{Q(x,y)}
\end{equation}
Therefore, we have
\begin{eqnarray}
\lim_{N\to\infty}\frac{\ln Z_{\mbox{\tiny ec}}(\beta)}{N}&=&
\sup_{\{\epsilon:~s_{X|Y}(\epsilon)\ge R_X\}}[s_{X|Y}(\epsilon)-R_X-\beta \epsilon]\\
&=&\left\{\begin{array}{ll}
\phi_X(\beta)-R_X & \beta < \beta_X\\
-\beta \epsilon_X & \beta \ge \beta_X\end{array}\right.
\end{eqnarray}
and 
\begin{equation}
\lim_{N\to\infty}\frac{\ln Z_{\mbox{\tiny ce}}(\beta)}{N}=
\left\{\begin{array}{ll}
\phi_Y(\beta)-R_Y & \beta < \beta_Y\\
-\beta \epsilon_Y & \beta \ge \beta_Y\end{array}\right.
\end{equation}
where 
\begin{eqnarray}
\phi_X(\beta)&=&\sum_yP(y)\ln\left[\sum_xP^\beta(x,y)\right],\\
\phi_Y(\beta)&=&\sum_xP(x)\ln\left[\sum_yP^\beta(x,y)\right],
\end{eqnarray}
$\epsilon_X$ is the solution to the equation $s_{X|Y}(\epsilon)=R_X$,
$\beta_X=s_{X|Y}'(\epsilon_X)$,
$\epsilon_Y$ is the solution to the equation $s_{Y|X}(\epsilon)=R_Y$,
and $\beta_Y=s_{Y|X}'(\epsilon_Y)$. Similarly,
\begin{eqnarray}
\lim_{N\to\infty}\frac{\ln Z_{\mbox{\tiny ee}}(\beta)}{N}&=&
\sup_{\{\epsilon:~s_{XY}(\epsilon)\ge R_X+R_Y\}}[s_{XY}(\epsilon)-R_X-R_Y-\beta
\epsilon]\\
&=&\left\{\begin{array}{ll}
\phi_{XY}(\beta)-R_X-R_Y & \beta < \beta_{XY}\\
-\beta \epsilon_{XY} & \beta \ge \beta_{XY}\end{array}\right.
\end{eqnarray}
where 
\begin{equation}
\phi_{XY}(\beta)=\ln\left[\sum_{x,y}P^\beta(x,y)\right],
\end{equation}
$\epsilon_{XY}$ is the solution to the equation $s_{XY}(\epsilon)=R_X+R_Y$,
and $\beta_{XY}=s_{XY}'(\epsilon_{XY})$.
Here, the phase diagram, in the three--dimensional space $(T,R_X,R_Y)$ is much more involved since
each one of the terms $Z_{\mbox{\tiny ee}}(\beta)$, $Z_{\mbox{\tiny
ce}}(\beta)$, $Z_{\mbox{\tiny ec}}(\beta)$ could be in two different phases,
and on top of that, one should check when $Z_{\mbox{\tiny cc}}(\beta)$
dominates. We will not delve into it any further here, but only note that for $\beta\le 1$
(which guarantees that all three erroneous partition
functions are in the paramagnetic phases), $Z_{\mbox{\tiny ee}}(\beta)$,
$Z_{\mbox{\tiny
ec}}(\beta)$, $Z_{\mbox{\tiny ce}}(\beta)$ and $Z_{\mbox{\tiny cc}}(\beta)$,
dominates, wherever $R_X+R_Y-\phi_{XY}(\beta)$, $R_X-\phi_X(\beta)$, 
$R_Y-\phi_Y(\beta)$, and
$\beta H(X,Y)$, is the smallest among all four functions, respectively.
In particular, we have the following conditions for
reliable communication (where $Z_{\mbox{\tiny cc}}(\beta)$ dominates):
\begin{eqnarray}
R_X&>&\beta H(X,Y)+\phi_X(\beta)\\
R_Y&>&\beta H(X,Y)+\phi_Y(\beta)\\
R_X+R_Y&>&\beta H(X,Y)+\phi_{XY}(\beta)
\end{eqnarray}
For $\beta=1$, this boils down to the well--known achievability region of SW
coding. Note that there are regions where either $Z_{\mbox{\tiny ec}}(\beta)$
or $Z_{\mbox{\tiny ce}}(\beta)$ dominate, which means that one of the sources
is decoded reliably, while the other one is not. As expected,
for $\beta=1$, $\by$ alone is decoded reliably within $\{(R_X,R_Y):~R_X <
H(X|Y),~R_Y > H(Y)\}$ and
$\bx$ alone is decoded reliably within $\{(R_X,R_Y):~R_Y <
H(Y|X),~R_X > H(X)\}$. 

\section{Performance Evaluation}

In this section, we provide an exact analysis of the error exponent,
associated with the symbol error probability of the finite--temperature decoder
(\ref{ft}). We then discuss some properties of the error exponent as a
function of $R$ and $\beta$ and present a phase diagram. Finally, we discuss 
some modifications and extensions.

For the sake of simplicity of the exposition, and without any essential loss
of generality, we will assume $\calX=\{0,1\}$ and evaluate the
expected\footnote{Expectation w.r.t.\ the random binning.} bit--error
rate (BER),
$P_b(R,\beta,N)=\mbox{Pr}\{\hat{x}_1\neq x_1\}$ (which is
the same as $\mbox{Pr}\{\hat{x}_i\neq x_i\}$ for every
$i=1,2,\ldots,N$ due to symmetry). that is,
\begin{eqnarray}
P_b(R,\beta,N)&=&\mbox{Pr}\left\{\sum_{\{\bx':~x_1'\ne x_1\}}P^\beta(\bx',\by)\cdot
I[f(\bx')=f(\bx)]\ge \right.\nonumber\\
& &\left.\sum_{\{\bx':~x_1'= x_1\}}P^\beta(\bx',\by)\cdot
I[f(\bx')=f(\bx)]\right\},
\end{eqnarray}
or more precisely, the error exponent associated with $P_b(R,\beta,N)$:
\begin{equation}
{\mathbb E}(R,\beta)\dfn\lim_{N\to\infty}\left[-\frac{\ln P_b(R,\beta,N)}{N}\right].
\end{equation}
For later use, we also define the following notation.
\begin{equation}
\epsilon(Q_{XY})\dfn\frac{1}{N}\ln P(\bx,\by)
=\sum_{(x,y)\in\calX\times\calY}Q_{XY}(x,y)\ln P_{XY}(x,y),
\end{equation}
where $Q_{XY}$ is understood here to be the joint empirical distribution of
$(\bx,\by)\in\calX^N\times\calY^N$. Similarly, for a generic $\bx'$, the 
corresponding auxiliary random variable will be denoted by $X'$, so that
the joint empirical distribution of
$(\bx',\by)$ will be denoted by $Q_{X'Y}$. In general, depending on the
context, $Q_{XY}$ and $Q_{X'Y}$ (or just $Q$)
may also denote generic joint distributions on $\calX\times\calY$, not
necessarily empirical distributions pertaining to sequences of finite length.
For a given $Q_{XY}$,
let us define
\begin{equation}
A(Q_{XY},R,\beta)\dfn \min_{Q_{X'|Y}}\left\{[R-H_Q(X'|Y)]_+:~\epsilon(Q_{X'Y})
+\frac{1}{\beta}[H_Q(X'|Y)-R]_+\ge \epsilon(Q_{XY})\right\},
\end{equation}
where $H_Q(X'|Y)$ is the conditional entropy of $X'$ given $Y$ associated with
$Q_{X'Y}$. Finally, define
\begin{equation}
E(R,\beta)=\min_{Q_{XY}}[D(Q_{XY}\|P)+A(Q_{XY},R,\beta)],
\end{equation}
where $D(Q_{XY}\|P)$ is the relative entropy (Kullback--Leibler
divergence) between $\{Q_{XY}(x,y)\}$ and $\{P(x,y)\}$, i.e.,
\begin{equation}
D(Q_{XY}\|P)=\sum_{x,y}Q_{XY}(x,y)\ln\frac{Q_{XY}(x,y)}{P(x,y)}.
\end{equation}
The following theorem presents our main result in this section.

\begin{theorem}
For the ensemble of random binning pertaining to SW codes, as described in 2.2,
\begin{equation}
{\mathbb E}(R,\beta)=E(R,\beta).
\end{equation}
\end{theorem}

\begin{figure}[ht]
\hspace*{2cm}\input{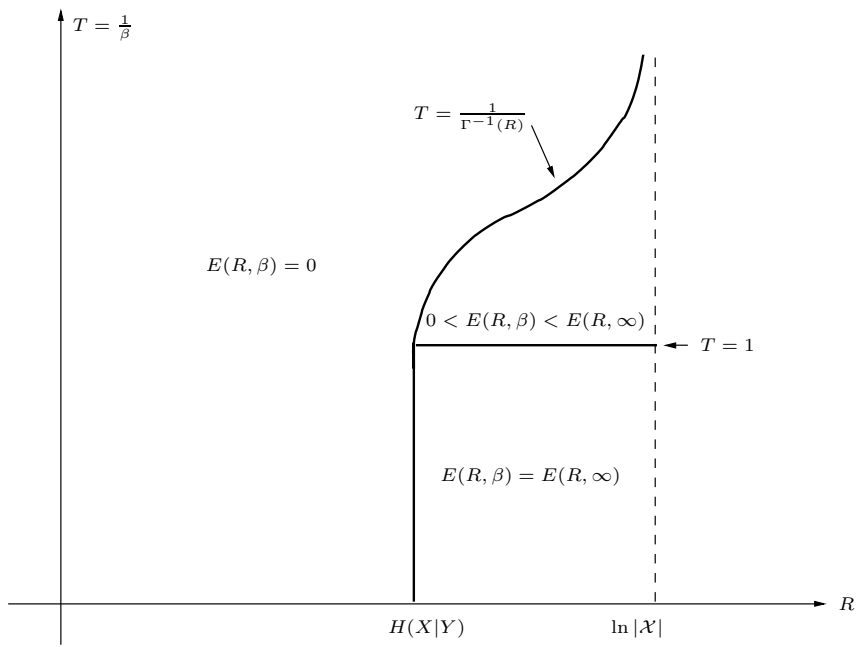}
\caption{\small Phase diagram of $E(R,1/T)$ in the plane of the decoding
temperature $T$ vs.\
the coding rate $R$.}
\label{fig2}
\end{figure}

It is easy to see that
$E(R,\beta)$ is non-decreasing both in $\beta$ and $R$. In particular,
\begin{equation}
E(R,\infty)=\min_{Q_{Y|X}}\left\{D(Q_{XY}\|P)+\min_{Q_{X'|Y}:~\epsilon(Q_{X'Y})\ge
\epsilon(Q_{XY})}[R-H_Q(X'|Y)]_+\right\},
\end{equation}
which agrees with the error exponent of the word error probability, as
expected. On the other hand, for $\beta=1$, the finite--temperature decoder
minimizes the BER and hence maximizes the exponent. Consequently, $E(R,\beta)$ must be a
constant for all $\beta\ge 1$, which is equal to $E(R,\infty)$.
It is also easy to verify that $E(R,\beta)$
vanishes for every
$\beta\le \Gamma^{-1}(R)$, i.e., beyond the paramagnetic--ferromagnetic
boundary curve.
Thus, $E(R,\beta)$ has three phases in the plane of $\beta$ vs.\ $R$:
(i) $\beta\le\Gamma^{-1}(R)$ or $R\le H(X|Y)$ (the union of the paramagnetic and
glassy phases of the posterior), where $E(R,\beta)=0$, (ii)
$R> H(X|Y)$ and $\beta\ge 1$, where $E(R,\beta)=E(R,\infty)$ (first
ferromagnetic
sub--phase), and (iii)
$R> H(X|Y)$ and $\Gamma^{-1}(R)\le \beta < 1$ (second ferromagnetic sub--phase),
where $E(R,\beta)<E(R,\infty)$
is monotonically non--decreasing both in $\beta$ and $R$.
The phase diagram of $E(R,\beta)$ is depicted
in Fig.\ \ref{fig2}. As can be seen, it is related to the phase diagram of the
finite--temperature partition function, but somewhat different. Here the
paramagnetic and the glassy
phases are united (in both of them $E(R,\beta)=0$), but the ferromagnetic
phase is subdivided into two new phases, as described above.

\noindent
{\it Remark 2.} In order to analyze the performance of other decoding metrics that
depend on $(\bx,\by)$ only via their joint type, one should simply replace the
definition
of $\epsilon(Q_{XY})$ by the corresponding metric, for example,
a mismatched metric $\epsilon(Q_{XY})=\sum_{x,y}
Q_{XY}(x,y)\ln \tilde{P}(x,y)$, or the minimum conditional
entropy metric $\epsilon(Q_{XY})=-H_Q(X|Y)$. Concerning the latter,
the phase diagram of the error exponent will be based on Fig.\
\ref{fig1u} in the same way that the phase diagram of Fig.\ \ref{fig2}
is based on the phase diagram of Fig.\ \ref{fig1}. Here, however, there is no
apparent subdivision of the ferromagnetic 
phase $E(R,\beta)> 0$ by the line $T=1$.
In other words, there are just two phases, $E(R,\beta)> 0$ and $E(R,\beta)=0$.

\vspace{0.1cm}

\noindent
{\it Proof of Theorem 1.}
The proof is similar to the proof of Theorem 1 in \cite{bid}.
For a given $(\bx,\by)\in\calX^N\times\calY^N$, and a given joint probability distribution
$Q_{X'Y}$ on
$\calX\times\calY$, let
$\Omega_1(Q_{X'Y})$ denote the number
of $\{\bx'\}$ within the bin of $\bx$, such that $x_1'\ne x_1$
and such that the empirical joint distribution with $\by$ is given by
$Q_{X'Y}$, that is
\begin{equation}
\Omega_1(Q_{X'Y})=\sum_{\bx':~x_1'=x_1}\calI\{(\bx',\by) \in
\calT(Q_{X'Y})\}\cdot\calI[f(\bx')=f(\bx)].
\end{equation}
For a given $(\bx,\by)$, the BER
is first calculated w.r.t.\
the randomness of the bins of codewords with $x_1'\ne x_1$, but
for a given binning of those with $x_1'=x_1$. 
We henceforth denote
$\calC_0=\{\bx':~\bx_1'=x_1,~f(\bx')=f\bx)\}$ and 
$\calC_1=\{\bx':~\bx_1'\ne x_1,~f(\bx')=f(\bx)\}$.

For a given
$\calC_0$ and $(\bx,\by)$, let
\begin{equation}
r\dfn \frac{1}{N}\ln\left[\sum_{\bx'\in\calC_0}P^\beta(\bx',\by)\right],
\end{equation}
and so, the BER becomes
$$\mbox{Pr}\left\{\sum_{\bx'\in\calC_1}P^\beta(\bx',\by)\ge
e^{Nr}\right\},$$ 
where it is kept in mind that $r$ is a function of
$\calC_0$ and $(\bx,\by)$. Now,
\begin{eqnarray}
\mbox{Pr}\left\{\sum_{\bx'\in\calC_1}P^\beta(\bx',\by)\ge e^{Nr}\right\}&=&
\mbox{Pr}\left\{\sum_{Q_{X'|Y}}\Omega_1(Q_{X'Y})e^{N\beta \epsilon(Q_{X'Y})}\ge
e^{Nr}\right\}\\
&\exe&\mbox{Pr}\left\{\max_{Q_{X'|Y}}\Omega_1(Q_{X'Y})e^{N\beta \epsilon(Q_{X'Y})}\ge
e^{Nr}\right\}\\
&=&\mbox{Pr}\bigcup_{Q_{X'|Y}}\left\{\Omega_1(Q_{X'Y})e^{N\beta \epsilon(Q_{X'Y})}\ge
e^{Nr}\right\}\\
&\exe&\sum_{Q_{X'|Y}}\mbox{Pr}\left\{\Omega_1(Q_{X'Y})e^{N\beta \epsilon(Q_{X'Y})}\ge
e^{Nr}\right\}\\
&\exe&\max_{Q_{X'|Y}}\mbox{Pr}\left\{\Omega_1(Q_{X'Y})\ge
e^{N[r-\beta \epsilon(Q_{X'Y})]}\right\},
\end{eqnarray}
Now, for a given $Q_{X'|Y}$, $\Omega_1(Q_{X'Y})$ is a binomial random variable with
exponentially $e^{NH_Q(X'|Y)}$ trials and probability of `success' 
$e^{-NR}$.
Thus, similarly as in \cite{bid} and \cite{CoN},
a standard large deviations analysis 
yields
\begin{equation}
\mbox{Pr}\left\{\Omega_1(Q_{X'Y})\ge
e^{N[r-\beta \epsilon(Q_{X'Y})]}\right\}\exe e^{-NE_0(r,\beta,R,Q_{X'Y})},
\end{equation}
where
\begin{eqnarray}
E_0(r,\beta,R,Q_{X'Y})&=&\left\{\begin{array}{ll}
[R-H_Q(X'|Y)]_+ & \beta \epsilon(Q_{X'Y})\ge s\\
0 & \beta \epsilon(Q_{X'Y})< r,~\beta \epsilon(Q_{X'Y})\ge r-H_Q(X'|Y)+R\\
\infty & \beta \epsilon(Q_{X'Y})< r,~\beta \epsilon(Q_{X'Y})<
r-H_Q(X'|Y)+R\end{array}\right.\nonumber\\
&=&\left\{\begin{array}{ll}
R-H_Q(X'|Y) & \beta \epsilon(Q_{X'Y})\ge r,~H_Q(X'|Y)\le R\\
0 & \beta \epsilon(Q_{X'Y})\ge r,~H_Q(X'|Y)\ge R\\
0 & \beta \epsilon(Q_{X'Y})< r,~\beta \epsilon(Q_{X'Y})\ge s-H_Q(X'|Y)+R\\
\infty & \beta \epsilon(Q_{X'Y})< r,~\beta \epsilon(Q_{X'Y})<
r-H_Q(X'|Y)+R\end{array}\right.\nonumber\\
&=&\left\{\begin{array}{ll}
R-H_Q(X'|Y) & \beta \epsilon(Q_{X'Y})\ge r,~H_Q(X'|Y)\le R\\
0 & \beta \epsilon(Q_{X'Y})\ge r-[H_Q(X'|Y)-R]_+,~H_Q(X'|Y)\ge R\\
\infty & \beta \epsilon(Q_{XY})< r-[H_Q(X'|Y)-R]_+\end{array}\right.\nonumber\\
&=&\left\{\begin{array}{ll}
[R-H_Q(X'|Y)]_+ & \beta \epsilon(Q_{X'Y})\ge r-[H_Q(X'|Y)-R]_+\\
\infty & \beta \epsilon(Q_{X'Y})< r-[H_Q(X'|Y)-R]_+\end{array}\right.
\end{eqnarray}
Therefore, $\max_{Q_{X'|Y}}\mbox{Pr}\{\Omega_1(Q_{X'Y})\ge
e^{N[r-\beta \epsilon(Q_{X'Y})]}\}$ decays according to
$$E_1(r,\beta,R,Q_Y)=\min_{Q_{X'|Y}}E_0(r,\beta,R,Q_{X'Y}),$$
which is given by
\begin{equation}
E_1(r,\beta,R,Q_Y)=\min\{[R-H_Q(X'|Y)]_+:~\beta \epsilon(Q_{X'Y})+[H_Q(X'|Y)-R]_+\ge r\}
\end{equation}
with the understanding that the minimum over an empty set is defined as
infinity. Finally, $P_b(R,\beta,N)$ is the expectation of $e^{-NE_1(r,\beta,R,Q_Y)}$
where the expectation is w.r.t.\ the randomness of the binning in $\calC_0$
and the randomness of $(\bX,\bY)$.
This expectation
will be taken in two steps: first, over the randomness of
the binning in $\calC_0$ while $\bx$ (the real transmitted
source vector)
and $\by$ are held fixed, and then
over the randomness of $\bX$ and $\bY$. Let $\bx$ and $\by$ be given
and let $\delta>0$ be arbitrarily small. Then,
\begin{eqnarray}
P_b(\bx,\by) &\dfn&
\bE[\exp\{-NE_1(r,\beta,R,Q_Y)\}|\bX=\bx,~\bY=\by]\nonumber\\
&=&\sum_r
P(r|\bX=\bx,~\bY=\by)\cdot
\exp\{-NE_1(r,\beta,R,Q_Y)\}\nonumber\\
&\le&\sum_i \mbox{Pr}\{i\delta\le
r<(i+1)\delta|\bX=\bx,~\bY=\by\}\cdot
\exp\{-NE_1(i\delta,\beta,R,Q_Y)\},
\end{eqnarray}
where $i$ ranges from $\frac{1}{N\delta}\ln P(\bx,\by)$ to some constant,
which is immaterial for our purposes.
Now,
\begin{eqnarray}
\label{ens}
e^{nr}&=&P^\beta(\bx,\by)+\sum_{\bx':~x_1'=x_1}P^\beta(\bx',\by)\cdot
\calI[f(\bx')=f(\bx)]\nonumber\\
&=& e^{N\beta \epsilon(Q_{XY})}+\sum_{Q_{X'|Y}}\Omega_0(Q_{X'Y})e^{N\beta
\epsilon(Q_{X'Y})},
\end{eqnarray}
where $Q_{XY}$ is the empirical distribution of $(\bx,\by)$ and
$\Omega_0(Q_{X'Y})$ is the number of codewords in $\calC_0\setminus\{\bx\}$ whose
joint empirical distribution
with $\by$ is $Q_{X'Y}$.
The first term in the second line of (\ref{ens})
is fixed at this stage. As for the second term, we have
(similarly as before):
\begin{equation}
\mbox{Pr}\left\{\sum_{Q_{X'|Y}}\Omega_0(Q_{X'Y})e^{N\beta \epsilon(Q_{X'Y})}\ge e^{nt}\right\}\exe
e^{-NE_1(t,\beta,R,Q_Y)}.
\end{equation}
On the other hand,
\begin{equation}
\mbox{Pr}\left\{\sum_{Q_{X'|Y}}\Omega_0(Q_{X'Y})e^{N\beta \epsilon(Q_{X'Y})}\le
e^{Nt}\right\}\exe
\mbox{Pr}\bigcap_{Q_{X'|Y}}\left\{\Omega_0(Q_{X'Y})\le e^{N[t-\beta \epsilon(Q_{X'Y})]}\right\}.
\end{equation}
Now, if there exists at least one $Q_{X'|Y}$ for which $R<
H_Q(X'|Y)$
and $H_Q(X'|Y)-R>t-\beta \epsilon(Q_{X'Y})$, then this $Q_{X'|Y}$ alone is responsible for a
double exponential decay of the probability of the event $\{\Omega_0(Q_{X'Y})\le
e^{N[t-\beta \epsilon(Q_{X'Y})]}\}$, let alone the intersection over all
$Q_{X'|Y}$.
On the other hand, if for every $Q_{X'|Y}$, either $R\ge
H_Q(X'|Y)$
or $H_Q(X'|Y)-R\le t-\beta \epsilon(Q_{X'Y})$, then we have an intersection of
polynomially many events whose probabilities all tend to unity. Thus, the
probability in question behaves exponentially like an indicator function of
the condition that for every $Q_{X'|Y}$, either $R\ge H_Q(X'|Y)$
or $H_Q(X'|Y)-R\le t-\beta \epsilon(Q_{X'Y})$, or equivalently,
\begin{equation}
\mbox{Pr}\left\{\sum_{Q_{X'Y}}\Omega_0(Q_{X'Y})e^{N\beta \epsilon(Q_{X'Y})}\le
e^{Nt}\right\}\exe
\calI\left\{R\ge\max_{Q_{X'|Y}}\{H_Q(X'|Y)-[t-\beta \epsilon(Q_{X'Y})]_+\}\right\}.
\end{equation}
Let us now find what is the minimum value of $t$ for which the value of
this indicator function is unity.
The condition is equivalent to
\begin{equation}
\max_{Q_{X'|Y}}\min_{0\le a\le 1}\{H_Q(X'|Y)-a[t-\beta \epsilon(Q_{X'Y})]\}\le R
\end{equation}
or:
\begin{equation}
\forall Q_{X'|Y}~\exists 0\le a\le 1:~~H_Q(X'|Y)-a[t-\beta \epsilon(Q_{X'Y})]\le R,
\end{equation}
which can also be written as
\begin{equation}
\forall Q_{X'|Y}~\exists 0\le a\le 1:~~t\ge
\beta \epsilon(Q_{XY})+\frac{H_Q(X'|Y)-R}{a}
\end{equation}
or equivalently,
\begin{eqnarray}
t&\ge&\max_{Q_{X'|Y}}\min_{0\le a\le
1}\left[\beta \epsilon(Q_{X'Y})+\frac{H_Q(X'|Y)-R}{a}\right]\\
&=&\max_{Q_{X'|Y}}\left[\beta \epsilon(Q_{X'Y})+\left\{\begin{array}{ll}
H_Q(X'|Y)-R &
H_Q(X'|Y)-R\ge 0\\ -\infty & H_Q(X'|Y)<R \end{array}\right.\right]\\
&=&\max_{\{Q_{X'|Y}:~R\le H_Q(X'|Y)\}}[\beta \epsilon(Q_{X'Y})+H_Q(X'|Y)]-R\\
&\dfn&r_0(Q_Y).
\end{eqnarray}
It is easy to check that $E_1(t,\beta,R,Q_Y)$ vanishes for $t\le r_0(Q_Y)$.
Thus, in summary, we have
\begin{equation}
\mbox{Pr}\left\{e^{nt}\le \sum_{Q_{X'|Y}}\Omega_0(Q_{X'Y})e^{N\beta \epsilon(Q_{X'Y})}\le
e^{n(t+\epsilon)}\right\}\exe\left\{\begin{array}{ll}
0 & t< r_0(Q_Y)-\epsilon\\
e^{-nE_1(t,\beta,R,Q_Y)} & t\ge r_0(Q_Y)\end{array}\right.
\end{equation}
Therefore, for a given $(\bx,\by)$, the expected error probability w.r.t.\
the randomness of the binning at $\calC_0$ yields
\begin{eqnarray}
P_e(\bx,\by)&=&\bE\{e^{-N[E_1(r,\beta,R,Q_Y)}|\bX=\bx,~\bY=\by\}\\
&\le& \sum_{i}\mbox{Pr}\left\{e^{Ni\delta}\le
\sum_{Q_{X'|Y}}\Omega_0(Q_{X'Y})e^{N\beta \epsilon(Q_{X'Y})}\le
e^{N(i+1)\delta)}\right\}\times\nonumber\\
& &\exp\{-NE_1(\max\{i\delta,\beta \epsilon(Q_{XY})\},\beta,R,Q_Y)\}\\
&\lexe& \sum_{i\ge r_0(Q_Y)/\delta}
\exp\{-NE_1(i\delta,\beta,R,Q_Y)\}\times\nonumber\\
& &\exp\{-NE_1(\max\{i\delta,\beta \epsilon(Q_{XY})\},\beta,R,Q_Y)\},
\end{eqnarray}
where the expression $\max\{i\delta,\beta \epsilon(Q_{XY})\}$ in the argument of
$E_1(\cdot,Q_Y)$ is due to the fact that
\begin{eqnarray}
r&=&\frac{1}{N}\ln\left[e^{N\beta
\epsilon(Q_{XY})}+\sum_{Q_{X'|Y}}\Omega_0(Q_{X'Y})e^{N\beta \epsilon(Q_{X'Y})}\right]\\
&\ge&\frac{1}{N}\ln\left[e^{N\beta \epsilon(Q_{XY})}+e^{Ni\delta}\right]\\
&\exe&\max\{i\delta,\beta \epsilon(Q_{XY})\}.
\end{eqnarray}
By using the fact that $\delta$ is arbitrarily small, we obtain
\begin{eqnarray}
P_e(\bx,\by)&\exe&\exp\{-NE_1(\max\{r_0(Q_Y),\beta
\epsilon(Q_{XY})\},\beta,R,Q_Y)\}\nonumber\\
&=&\exp\{-N\max\{E_1(r_0(Q_Y),\beta,R,Q_Y),E_1(\beta
\epsilon(Q_{XY}),\beta,R,Q_Y)\}\nonumber\\
&=&\exp\{-NE_1(\beta
\epsilon(Q_{XY}),\beta,R,Q_Y)\}
\end{eqnarray}
since the dominant contribution to the sum over $i$ is due to the term
$i=r_0(Q_Y)/\delta$ (by the non--decreasing monotonicity of the function
$E_1(\cdot,Q_Y)$).
After averaging w.r.t.\ $(\bX,\bY)$, we obtain
\begin{eqnarray}
E(R,\beta)&=&\min_{Q_{XY}}\{D(Q_{XY}\|P)+E_1(\beta
\epsilon(Q_{XY}),\beta,R,Q_Y)\}\\
&=&\min_{Q_{XY}}\{D(Q_{XY}\|P)+A(Q_{XY},R,\beta)\},
\end{eqnarray}
completing the proof of Theorem 1.


\end{document}